\documentstyle[12pt]{article}

\def\be{\begin{equation}}
\def\ee{\end{equation}}
\def\bea{\begin{eqnarray}}
\def\eea{\end{eqnarray}}

\def\beq{\begin{equation}}   \def\eeq{\end{equation}}
\def\bea{\begin{eqnarray}}   \def\eea{\end{eqnarray}}

\begin{document}
\begin{flushright}
UND-HEP-02-BIG\hspace*{.2em}06\\
TTP02-07 \\
hep-ph/0206225 \\
June 2002 \\
\end{flushright}


\vspace{.3cm}
\begin{center} \Large 
{\bf 
{THE VIETRI CODICES}}
\footnote{Summary talk given at {\em HQ\&L 2002}, 
International Workshop on Heavy Quarks and Leptons, 
Vietri sul Mare, Salerno, Italy, 
May 27 - June 1, 2002}
\\
\end{center}
\vspace*{.3cm}
\begin{center} {\Large 
I. I. Bigi }\\ 
\vspace{.4cm}
{\normalsize 
{\it Physics Dept.,
Univ. of Notre Dame du
Lac, Notre Dame, IN 46556, U.S.A.}
\footnote{Permanent address} \\
and \\
{\it Institut f\" ur Theoret. Teilchenphysik, 
Universit\" at Karlsruhe, D-76128 Karlsruhe, 
FR Germany} }
\\
\vspace{.3cm}
{\it e-mail address: bigi.1@nd.edu } 
\vspace*{0.4cm}

{\Large{\bf Abstract}}
\end{center}
Studying heavy flavour physics is driven by multiple 
motivations: to probe our theoretical control over QCD;  
enhance our information on gluon and sea quark structure 
functions; 
use heavy flavour production as signal for the onset of the quark-gluon
plasma;  revisit light flavour spectroscopy through the final states 
in heavy flavour decays; 
extract fundamental quantities like CKM and MNS parameters; 
search for new Physics. 
The last two years have witnessed tremendous amplification in our 
knowledge of heavy flavour dynamics. We know the elements 
of the CKM matrix with significantly improved reliability; a new 
door has been opened by the observation of 
$K^+ \to \pi ^+ \nu \bar \nu$; with {\em direct} CP violation 
being established in $K_L$ decays and the expected huge CP asymmetry 
been observed in $B_d(t) \to \psi K_S$ the CKM description has 
attained the level of a {\em tested theory} of CP violation. 
Yet more than ever the Standard Model is viewed as incomplete: 
apart 
from the peculiar pattern in mass related parameters of the 
fundamental fermions there is more direct evidence in the 
signals for neutrino
oscillations, the strong CP problem 
and the baryon number of the universe. 
A worldwide, interrelated and comprehensive program has been 
developed for heavy quark and lepton physics that will provide 
essential impetus towards a deeper understanding of nature's 
grand design.


\tableofcontents 
\section{Reflections}
\label{INTRO}

Fifty-five years ago two events happened that typify the study 
of fundamental dynamics central to this conference: in 
1947 Rochester and Butler published a paper 
\cite{BUTLER} on the observation of kaons 
decaying into two pions; in the same year Purcell and Ramsey 
\cite{PURCELL} started 
the journey towards measuring an electric dipole moment for neutrons. 
While we have learnt everything that can be learnt 
from $K\to \pi \pi$, this is far from true for the journey to $d_N$.

There are three reasons why I have chosen the title "Vietri Codices" 
for this summary: 
\begin{itemize}
\item 
Written information was originally stored on rolls. Yet as 
Callimachus  
working at the ancient library of Alexandria in the 
third century B.C. once remarked: 
"A big roll is a big nuissance." The form of the {\em codex} or 
book was developed as a first form of 
R(andom)A(ccess)M(emory). A  
summary talk can provide you with multiple, 
though not random access. 
\item 
Leonardo da Vinci had the habit of 
writing `parity inverted'. A normal mortal therefore has to 
read his writings -- like the Leicester Codex containing his 
scientific observations -- through a mirror
\cite{LEO}. The symbolism for 
the topics of this conference is obvious. 
\item 
The celebrated Dresden Codex with symbols from the Mayas 
must contain profound elements of their knowledge, presumably 
about astronomy. Unfortunately one has not found its dictionary. 
One hopes to infer its contents through uncovering correlations 
between its symbols and celestial events. We are facing a similar 
interpretative challenge; we have actually 
(at least) two Dresden codices in 
front of us, namely the CKM and MNS matrices. We can be certain they 
contain profound messages about truly fundamental dynamics -- 
alas it is highly coded with even the synthax  
mysterious. The analogy is even visual: looking at a page from 
the Dresden Codex \cite{DRESDEN} you see 
groupings of three dots together with symbols similar to matrices 
and mass  spectra of neutrinos. 
\end{itemize}

This conference focusses on heavy quarks and leptons. At first it might 
seem they have little in common beyond charge 
quantization: quarks are subject to
the strong interactions and are actually confined. You often hear a lot
of  moaning that this introduces all kinds of 
unpleasant hadronization effects. Whenever feeling thus challenged one
can remember the following eternal truths:  there is no royal way to
knowledge; hardship builds  character and there is salvation
at the very end. You might find little solace in such advice thinking it
merely reflects the  teachings of my employer. However I insist you
should view hadronization  as a true blessing even if well disguised. For
without hadronization  there would be no $K^0 - \bar K^0$ oscillations
providing such a  glorious lab to reveal indirect CP violation, and direct
CP violation  could not manifest itself through the observables 
$\eta _{+-} \neq \eta _{00}$. Likewise there could be no 
$B^0 - \bar B^0$ oscillations and no huge CP violation in 
$\Delta B =2$ transitions; the latter could not serve as a highly 
sensitive probe for New Physics. More specifically, the formation 
of the $K_L$ and $K_S$ mass eigenstates with the former being just 
above the 3 pion threshold presents us with two powerful tools: 
it provides a phase space suppression of the CP {\em allowed} 
$K_L \to 3\pi$ modes by a factor of $\sim 500$ relative to the 
CP forbidden $K_L \to \pi \pi$ channels, and at the same time it 
allows CP violation to manifest itself through the existence of a 
transition rather than an asymmetry.

One can understand also on 
very general grounds that the
strong interactions  should be viewed as the hero rather than the villain
in the  story of CP violation: 
since the latter can arise through a complex phase only 
due to CPT invariance, one needs two different, yet coherent 
amplitudes for an asymmetry to emerge; hadronization provides such a 
second amplitude -- $K^0 \Rightarrow \bar K^0 \to f$ in addition 
to $K^0 \to f$ -- and `cools' the degrees of freedom to increase 
considerably the coherence and thus interferebility of the two 
amplitudes.  Later I will tell you about an example why we
should be very  careful in what we wish.

The dynamics of heavy flavour hadrons reflects the interplay of strong 
and electroweak forces. Despite our problems in establishing
computational control over QCD I would characterize the situation as 
follows: "QCD is the only thing -- $SU(2)_L \times U(1)$ not even  
the greatest thing!" I.e., there is no alternative to QCD among 
local quantum field theories. Therefore I find it quite inappropriate 
to state we are {\em testing} QCD: for it would imply that a failure 
would lead to QCD being discarded; this, however, is not going to
happen. The correct statement is that we are {\em probing} QCD to 
understand its inner workings. A
failure in such a probe only means that we have to learn from our
mistakes and adjust our calculational tools.

There are five classes of possible lessons, of which the first three
refer to QCD and the remaining two to the electroweak  sector: 
\begin{itemize}
\item 
Refining theoretical technologies for QCD, namely perturbative QCD, 
chiral perturbation theory, $1/m_Q$ expansions and lattice QCD;   
\item    
extracting gluon and sea structure functions of the nucleon 
\cite{PAUL};  
\item 
doing our homework on heavy flavour production to interprete 
heavy ion collisions \cite{GOTT}; 
\item 
determining fundamental quantities like the CKM and MNS parameters 
and fermion masses; 
\item 
searching for New Physics -- the Great Prize \cite{MARCH,MARTINELLI} . 
\end{itemize}
The remainder of my talk will be organized as follows: in 
Sect.\ref{PROD} I will summarize the presentations on heavy 
flavour production; in Sect.\ref{DECAY} I will deal 
with the weak decays of strange, charm and beauty hadrons; after 
sketching the CP landscape in Sect.\ref{LAND} I will describe the 
unreasonable success of the CKM theory in Sect.\ref{UNREAS} before turning
to searches for  New Physics in Sect.\ref{NEW}; after addressing recent
results  presented here on 
the lepton sector in Sect.\ref{LEP} I will offer some concluding 
remarks.

\section{Production of Heavy Flavours}
\label{PROD}

Heavy flavour production can be categorized by the numbers of hadrons 
in the initial state: zero hadrons for $e^+e^-$ annihilation, one hadron 
for deep inelastic lepton-nucleon scattering, $(1+\alpha )$ and
two hadrons  for photo- and hadro-production, respectively. Obviously 
the theoretical challenge increases with the number of hadrons.

No excuses would be accepted for hadroproduction of top quarks 
(top {\em hadrons} do not form \cite{DOK}) -- and none are needed
\cite{NASON}.

Hadroproduction of beauty hadrons provides an intriguing tale
\cite{NASON}. Both CDF and D0 have reported beauty production rates 
exceeding predictions by factors of two to three. Most recently 
CDF has stated an excess factor of $2.9 \pm 0.2 \pm 0.4$ 
\cite{CDF}. This is hard
to  swallow considering the large beauty mass and that one is dealing with
central production of beauty. Beauty 
fragmentation might hold the key to the puzzle. Using a harder
fragmentation function than the Peterson et al. implementation the 
authors of Ref.\cite{NASON2} find a reduced factor of 
$1.7 \pm 0.5|_{th} \pm 0.5|_{exp}$, i.e. hardly a significant excess.
Future data  will shed light on it.

We should not declare victory yet. As reported by Gladilin \cite{GLAD}, 
beauty production in $ep$ collisions "typically" exceeds predictions.

A great deal can be 
learnt from charm production in $\nu$ scattering \cite{MIGL}: $|V(cs)| \&
|V(cd)|$  can be extracted, the 
{\em absolute} value of $BR_{SL}(\Lambda _c)$
be  determined without undue reliance on a model etc. I want to state a 
caveat though concerning the charm quark mass. It represents a much
more 
subtle object than a parameter in a quark-parton 
model. With quarks being 
confined, there is no `natural' definition of a quark mass; there is 
actually an infinity of possible definitions of quark masses in a 
quantum field theory taken at different scales. I suspect that the charm 
quark mass receives significant nonperturbative contributions that are 
specific to the production process. The tools are there to analyze this 
issue, yet to my knowledge it has not been done. However as long as 
one uses it pragmatically to parametrize the threshold behaviour in
a  {\em given} reaction without any claim to have determined a 
fundamental quantity, I would not object to it.

\section{Weak Decays of Heavy Flavour Hadrons}
\label{DECAY}

\subsection{Strange decays}

The theoretical tool of choice here is chiral perturbation theory. 
It is probed for its intrinsic reason, to extract 
$|V(us)|$ and to gain better theoretical control over 
various CP asymmetries \cite{COL}.

$|V(us)|$ is extracted from $K_{e3}$ decays. The `classic' analysis 
by Leutwyler \& Roos from 1984 \cite{ROOS} yields 
\beq 
|V(us)|_{K_{e3}} = 0.2196 \pm 0.0023 
\label{LR} 
\eeq
Weak Universality -- one of the fundamental features of the 
SM we understand -- requires 
\beq 
|V(ud)|^2 + |V(us)|^2 + |V(ub)|^2 = 1
\label{WU}
\eeq
Using the PDG 2000 value for $|V(ud)|$ 
one infers from Eq.(\ref{WU}) a
higher value  ($V(ub)$ is numerically insignificant):  
\beq 
|V(us)|_{unit} = 0.2287 \pm 0.0034 \; ; 
\label{WUVUS}
\eeq
i.e., the value in Eq.(\ref{LR}) does not satisfy the unitarity 
requirement. The elements of this argument have of course to be 
re-analyzed: 
\begin{itemize}
\item 
Updating the analysis of Leutwyler \& Roos Cirigliano {\em et al.} 
found almost no numerical change \cite{CIRI}
\beq 
|V(us)|_{K_{e3}} = 0.2201 \pm 0.0013|_{\Delta \Gamma} 
\pm 0.0008|_{\Delta \lambda _+} \pm 
0.0019|_{\Delta f_+(0)}
\eeq
\item 
A new measurement by E 865 finds the very preliminary value 
BR$(K_{e3}) = (5.13 \pm 0.02 \pm 0.08 \pm 0.04)\% $ 
\cite{DPF02}
to be 
compared to the PDG 2000 value of $(4.82 \pm 0.06)\%$. 
It would suggest $|V(us)|_{K_{e3}} = 0.2271 \pm 0.0024$, 
in full agreement with Eq.(\ref{WUVUS}). 
\item 
PDQ 2002 has adopted the following values as directly 
determined \cite{KLEIN}: 
\beq
|V(us)|_{K_{e3}} = 0.2196 \pm 0.0026 \; 
, \; 
|V(ud)| = 0.9734 \pm 0.0008
\label{PDG02} 
\eeq
where the latter represents an average over data on mirror nuclei and
free neutrons. 
\item 
Taking the {\em central} values of Eq.(\ref{PDG02}) one finds a 
2.7 $\sigma$ deficit for Eq.(\ref{WU}); on the other hand an overall 
fit {\em with weak universality built in} yields the ranges 
\beq 
|V(us)|_{unit} = 0.219 \div 0.226 \; , \; 
|V(ud)|_{unit} = 0.9741 \div 0.9756 \; ; 
\eeq
$|V(ud)|_{unit}$ is a touch high viz. Eq.(\ref{PDG02}), 
yet not conclusively so. 
\end{itemize} 
I think that clarification of this complex issue both requires and 
deserves considerable efforts on the experimental and  
theoretical side based on its intrinsic interest as well as a 
case study of indirect searches for New Physics. Personally I view it 
as highly unlikely that Eq.(\ref{WU}) could be violated by as much 
as 1 \%: for I believe that such a violation of weak universality should 
lead to observable effects in the electric dipole moment of neutrons.

Both the modulus and imaginary part of $V(td)$ can be determined 
in $K\to \pi \nu \bar \nu$ decays \cite{ISI,KETT}. To be more
specific:  The SM predicts for $K^+ \to
\pi ^+ \nu \bar \nu$ and 
$K_L \to \pi ^0 \nu \bar \nu$ branching ratios of 
$(7.2 \pm 2.1)\cdot 10^{-11}$ and $(2.6 \pm 1.2)\cdot 10^{-11}$, 
respectively. The large range in the two predictions reflects our 
ignorance about the value of $V(td)$ and shows why measurements 
of these two branching ratios will significantly enhance our knowledge; 
for the intrinsic uncertainty in the latter is estimated to be  around 2
\% and in the former around 7 \% (mostly due to ignorance  concerning the
charm quark mass). E 787 has seen two events, which  corresponds to a
branching ratio 
$(1.57 ^{+1.75}_{-0.82})\cdot 10^{-10}$, i.e. about double the SM 
expectation, yet not significantly different; on the other hand 
there is a $0.02\%$ probability for the background to generate 
two events. 
The successor experiment E 949 expects to see $\sim 5 - 10$
SM 
$K^+ \to \pi ^+ \nu \bar \nu$ 
events over the next two years or so to be followed by CKM aiming 
at about 100 SM events after 2008. The neutral mode 
$K_L \to \pi ^0 \nu \bar \nu$ has not been seen yet; KOPIO hopes to 
collect $\sim 50$ SM events after 2006. These are certainly 
challenging experiments, yet central in our quest to determine 
fundamental quantities and search for New Physics.

Two more points: 
\begin{itemize}
\item 
From NA48 data on $K_L \to \pi ^0 \gamma \gamma$ one can deduce that 
the CP conserving contribution to $K_L \to \pi ^0 e^+e^-$ is
insignificant \cite{WANKE}. 
\item 
It seems to me that the prize for the most improved player should 
go to KLOE which has begun to produce intriguing data 
\cite{DELUCIA}. I took note 
of their measurement of the relative rates for the radiative 
transitions $\phi \to \eta ^{\prime}\gamma$ vs. 
$\phi \to \eta \gamma$ where they find no evidence for a $gg$ component 
in the $\eta ^{\prime}$ state. I remember that at a place long ago and 
far away MARKIII analyzing the analogous decays 
$\psi \to \eta ^{\prime}\gamma$ vs. 
$\psi \to \eta \gamma$ arrived at different conclusions.

\end{itemize}

\subsection{Charm decays}

The width for $D^* \to D \pi$ has been measured by CLEO 
\cite{CLEO}; the value extracted for the strong 
$D^*D\pi$ coupling appears to be significantly higher 
than what can be accommodated with predictions based on light cone
sum rules \cite{KOD}. Since I feel those predictions cannot be 
dismissed easily, I advocate further experimental scrutiny.

The $1/m_Q$ expansion provides a 
decent semi-quantitative description of the lifetime ratios for 
charm hadrons; this could not be counted upon since the charm quark 
mass does not exceed ordinary hadronic scales by a large amount. 
Yet the $\Xi _c^+$ lifetime seems to be quite a bit longer 
than predicted \cite{BCP,GUB}; on the other hand the 
$\Xi_c^0$ lifetime
for which  there is a new preliminary number from FOCUS \cite{BIANCO}
\beq  
\tau (\Xi_c^0) = 118 ^{+14}_{-12} \pm 5 \; \; {\rm fs} 
\eeq 
seems to be in line with expectations.

Buccella described a phenomenological treatment of nonleptonic 
two-body modes of charm mesons on the Cabibbo allowed and once 
suppressed level \cite{BUCCELLA}. One has to be realistic in one's 
expectations concerning the accuracy of this ansatz. It is 
nontrivial that one gets a decent overall description without having 
to introduce too many `epicycles' in the form of resonances and other 
final state interactions. The most important lesson one wants to learn 
from such an exercise is to find out which channels exhibit the 
strongest final state interactions that would allow direct CP 
violation to surface there.

Miranda made a good case \cite{MIRANDA} that the final state in the 
decays of charm hadrons represent a novel and promising lab 
to study light flavour spectrocopy since one is dealing with 
exclusive states of well-known overall quantum numbers.  At
the same time some caveats have to be kept in mind like that  a
Breit-Wigner ansatz is an approximation of varying accuracy; in 
particular for scalar di-meson resonances a lot of information  exists 
from data that a Breit-Wigner ansatz provides in  general a poor
description \cite{MEISS}. More work is clearly needed  there, and 
theorists should spend some quality  time on this new
frontier.

Considerable information has been accumulated on the decay 
constant $f_{D_s}$ extracted from $D_s \to \tau \nu$ and 
$D_s \to \mu \nu$ \cite{MARI}. Experimental extractions range 
from about 200 MeV to above 300 MeV. The most recent and partially 
unquenched lattice results yield $240^{+30}_{-25}$ MeV.

In 1993 I had stated at a tau-charm factory workshop in Spain that 
the "tau-charm factory is the QCD machine for the '90's" 
\cite{MARBELLA}. I was 
wrong by just ten years, for it seems very likely now that Cornell 
can realize CLEO-c taking data in the charm region over the next 
three or so years.

As I will sketch later, there is a very considerable potential 
for New Physics to manifest itself in charm hadron transitions as  
can be studied at $B$ factories -- a point I will return to.

\subsection{Beauty decays}

The decays of beauty hadrons exhibit a rich and multilayered 
CKM phenomenology with interplay of all three families. 
With BABAR already exceeding its design luminosity of 
$3\cdot 10^{33} s^{-1}cm^{-2}$ and BELLE getting close to its 
even more ambitious 
design value of $10^{34} s^{-1}cm^{-2}$ data are 
accumulated in huge streams and will continue to do so 
\cite{WANG,ROBERTSON,YAMA}. This 
will allow us to conduct measurements in the foreseeable future 
that a few years ago we would not have dared to even contemplate.

BELLE has seen the new `colour suppressed' channels 
$B_d \to D^0 +\pi^0/\eta /\omega$ and also the unconventional 
mode $B^{\pm} \to p \bar p K^{\pm}$ with a branching ratio of 
$(4.3 ^{+1.1}_{-0.9} \pm 0.5)\cdot 10^{-6}$ \cite{WANG}.

We have a well-stocked chest of theoretical technologies of 
varying complexity and range of applicability: 
(i) Heavy quark expansions 
\cite{URI}, (ii) HQET, (iii) lattice QCD, (iv) QCD factorization 
\cite{NEUBERT} and (v) chiral dynamics \cite{MEISS}. They allow us 
to make predictions with a reliability, accuracy and breadth that 
ten years ago would have seemed merely wishful thinking. In 
semileptonic and radiative $B$ decays one can give detailed error 
budgets; a well defined kinetic $b$ quark mass can be extracted from 
$\Upsilon (4S) \to b \bar b$ -- 
$m_b^{kin}(1\; {\rm GeV}) = 4.57 \pm 0.08$ GeV 
\cite{MEL,HOANG,SIGNER} -- and from the 
moments in semileptonic $B$ decays -- 
$m_b^{kin}(1\; {\rm GeV}) = 4.65 \pm 0.10$ GeV \cite{URI}. It provides 
a highly remarkable test that these two methods, which are so 
different in both their experimental and theoretical aspects,  
yield results in such good agreement! This supports the 
statement that we have extracted $|V(cb)|$ with about 5 \% theoretical 
uncertainty \cite{URI}: 
\beq 
|V(cb)| = 0.0412 \cdot \left( 
1\pm 0.015_{\rm pert} \pm 0.01_{m_b} \pm 0.012 
\pm 0.012
\right)  \cdot \left[ 
\frac{{\rm BR}_{SL}(B)}{0.105}
\right] ^{\frac{1}{2}}
\eeq 
and that the latter can be reduced further already in the near 
future \cite{URI}. This state of maturity and interconnectedness is well
characterized by Uraltsev's statement "ill-defined parameters 
lead to observable problems".

Neubert \cite{NEUBERT} presented 
extensive tables with QCD based predictions on two-body modes of 
$B$ mesons; there appears no longer a 
"$\eta /\eta ^{\prime}$ problem".  
It was emphasized that these nonleptonic 
decays favour $\phi_3 [\gamma] > 90^o$, whereas 
$\Delta B = 2$ transitions -- 
$\Delta m(B_d)$ and the bound on $B_s - \bar B_s$ oscillations -- 
yield $\phi_3 [\gamma] < 90^o$.

A promising way to extract $\phi_2 [\alpha ]$ is from studying 
the evolution of $B_d(t) \to \rho \pi$ \cite{ART}. 
In the data there will be contributions from $B \to \sigma \pi$ 
as well; this complication can be dealt with \cite{MEISS}, 
although it presumably increases the level of statistics 
needed. Yet when aiming at an accuracy of a few \% , one has to 
evaluate very carefully whether the resonance structures are 
adequately described \cite{MEISS}.

\section{CP Violation -- The Landscape}
\label{LAND}

CP violation was discovered 1964 through the decay 
$K_L \to \pi ^+ \pi ^-$ -- causing considerable consternation 
among theorists \cite{LANDAU}. 
Till 1999, i.e. for 35 years, CP violation 
could be described by a {\em single} non-vanishing real number -- 
namely the phase between the quantities $M_{12}$ and 
$\Gamma _{12}$ in the $K^0 - \bar K^0$ mass matrix -- 
in face of a large body of data. {\em Direct} CP violation 
has been unequivocally established in 1999. In the summer of 
2001 peaceful coexistence has been achieved between the data of 
NA48 and KTeV with a new world average \cite{EPSPRIME}: 
\beq 
\langle \epsilon ^{\prime}/\epsilon _K \rangle = 
(1.72 \pm 0.18) \cdot 10^{-3} 
\eeq 
Quoting the result in this way does not do justice to the experimental 
achievement, since $\epsilon _K$ is a very small number itself. 
The sensitivity achieved and the control over systematic uncertainties 
established becomes more obvious when quoted in terms of actual 
widths:
\beq 
\frac{\Gamma (K^0 \to \pi ^+ \pi ^-) - 
\Gamma (\bar K^0 \to \pi ^+ \pi ^-)}
{\Gamma (K^0 \to \pi ^+ \pi ^-) + 
\Gamma (\bar K^0 \to \pi ^+ \pi ^-)} = 
(5.7 \pm 0.6) \cdot 10^{-6} \; !
\eeq
This represents a discovery of the very first rank -- 
{\em no} matter what theory does or does not say. We can take pride 
in this achievement. The two groups deserve our respect; they 
have certainly earned my admiration.

It had been predicted already 1980 that the CKM description implies 
large CP asymmetries in several classes of $B$ decays involving 
$B_d - \bar B_d$ oscillations, most
notably  in $B_d \to \psi K_S$ \cite{CS,BS}. The existence of a huge CP
asymmetry  in $B_d \to \psi K_S$ has been firmly established in 2001 by
BELLE and  BABAR; this spring they have presented
updates that agree very  nicely \cite{BABAR1}: 
\beq 
{\rm sin}2\phi _1[\beta] =  
\left\{ 
\begin{array}{l}
0.75 \pm 0.09 \pm 0.04 \; \; \; \; {\rm BABAR} \\
0.82 \pm 0.12 \pm 0.05 \; \; \; \; {\rm BELLE}
\label{PHIDATA}
\end{array}  
\right.
\eeq
After the meeting both groups have 
further updated their results \cite{BABAR2,YAMA}:
\beq 
{\rm sin}2\phi _1[\beta] =  
\left\{ 
\begin{array}{l}
0.741 \pm 0.067 \pm 0.033 \; \; \; \; {\rm BABAR} \\
0.719 \pm 0.074 \pm 0.035 \; \; \; \; {\rm BELLE}
\label{PHIDATA}
\end{array}  
\right.
\eeq
We can conclude 
that the CP asymmetry in $B_d \to \psi K_S$ is there for sure and it is 
huge -- as expected! For intellectual interest one can add that 
the measurements, which are based on EPR correlations \cite{EPR},  
illustrate nicely that CP violation is coupled with 
T violation \cite{KRAKOW}.

\section{The Unreasonable Success of the CKM Description}
\label{UNREAS}

In 1998, i.e. {\em before} $\epsilon ^{\prime} \neq 0$ and CP violation 
in $B_d$ decays were established, the status of the CKM description 
could be sketched as follows: 
$\Delta m_K$, $\epsilon _K$ and $\Delta m(B_d)$ 
could be reproduced; $\epsilon ^{\prime}/\epsilon _K \leq 10^{-3}$ 
was widely advocated except for some heretics \cite{TRIESTE,BERT}; 
concerning the CP asymmetries in $B$ decays it was stated that 
some have to be of order unity with no "plausible" deniability; 
in the early '90's -- i.e. before the discovery of top 
quarks -- this was specified to predicting 
sin$2\phi_1[\beta] \sim 0.6 - 0.7$ if today's estimates of 
$f_B$ are used \cite{BEFORETOP}.  
In '98 courageous souls gave a prediction of 
sin$2\phi_1[\beta] \sim 0.72 \pm 0.07$ 
\cite{LACKER}.

It is indeed true that large fractions of $\Delta m_K$, 
$\epsilon _K$ and $\Delta m_B$ and even most of 
$\epsilon ^{\prime}$ could be due to New Physics; constraints 
from data thus translate into `broad' bands in plots of the 
unitarity triangle. Yet such a statement seemingly reflecting 
facts misses the real point! One has to keep in mind that 
the dimensional quantities describing the weak observables 
span several orders of magnitude in energy units, 
namely the range 
${\cal O}(10^{-15}) \div {\cal O}(10^{-9})$ MeV. It is
highly remarkable that  the CKM description can always get to within a
factor of two  or three -- in particular with numerical values in its
parameters  and the fermion masses that {\em a priori} would have seemed 
to represent frivolous choices, like $m_t \simeq 180$ GeV. 
And it appears right on the mark for sin$2\phi_1$! Hence I 
conclude that the CKM description is no longer a mere ansatz, but 
a tested theory; its forces are with us to stay. 
En passant we have learnt that when complex phases surface they 
can be large. 
An aside might be allowed here. 
A CP odd quantity depends also on the
$sin$  (and $cos$) of the three CKM angles in addition to the complex 
phase. While the latter is large, the former are small or even 
tiny -- something that can be understood in the context of theories 
with extra dimensions \cite{MARCH} -- allowing CP violation a 
la CKM to be generated perturbatively \cite{DENT}. A large 
CP {\em asymmetry} can arise when also the decay rate is 
suppressed by small CKM parameters as it happens in $B$ decays.

Yet this new and spectacular success of the SM does not resolve 
any of its mysteries -- why are there families, why three, 
what is the origin of the peculiar pattern in the quark mass matrices 
-- they actually deepen them.  
Consider the structure of the CKM matrix: 
\beq 
|V_{CKM}| \sim 
\left( 
\begin{array}{ccc} 
1 & {\cal O}(\lambda ) & {\cal O}(\lambda ^3) \\ 
{\cal O}(\lambda ) & 1 & {\cal O}(\lambda ^2) \\ 
{\cal O}(\lambda ^3) & {\cal O}(\lambda ^2) & 1 
\end {array} 
\right) 
, \; \lambda = {\rm sin}\theta _C
\label{DRES}
\eeq
There has to be fundamental information encoded in this 
hierarchical pattern. 
The situation can be categorized by saying "we know so much, yet 
understand so little!" I.e., it re-emphasizes that the SM is 
incomplete, that New Physics must exist. Theories with extra dimensions 
might provide an answer \cite{MARCH}.

\section{Searching for New Physics}
\label{NEW}

\subsection{The `King Kong' scenario}

$\Delta S=1,2$ dynamics have provided several examples of revealing the 
intervention of features that represented New Physics 
{\em at that time}; 
it thus has been instrumental in the evolution of the SM. This happened 
through the observation of `qualitative' discrepancies; i.e., 
rates that were expected to vanish did not, or rates were 
found to be smaller than expected by several orders of magnitude. 
Such an indirect search for New Physics can be characterised as a 
`King Kong' scenario: "One might be unlikely to encounter King 
Kong; yet once it happens there can be no doubt that one has 
come across something extra-ordinary". Such a situation 
can be realized again in different ways: 
\begin{itemize}
\item 
Dedicated searches for EDM's of neutrons, atoms and molecules are 
a definite must as emphasized by Pospelow \cite{POSP} -- no excuses are 
acceptable. For one should keep in mind that they are so tiny in the 
SM for reasons very specific to the CKM implementation of T violation.  
\item 
Searching for a transverse polarization of muons in 
$K^+ \to \mu ^+ \nu \pi ^0$ is a promising way to uncover the 
intervention of Higgs based CP violation. 
\item 
There exists a large literature on $D^0 - \bar D^0$ oscillations 
with predictions for observables covering several orders of 
magnitude. That does not mean that they are all equally credible, 
though. A systematic analysis has been given now based on the 
operator product expansion expressing $x = \Delta m_D/\Gamma$ and 
$y=\Delta \Gamma/2\Gamma$ in powers of $1/m_c$, $m_s$ and KM factors. 
One finds \cite{DOSC} $x$, $y$ $\sim {\cal O}(10^{-3})$ with the 
prospects for reducing the uncertainties rather slim; one should also 
note that $y$ is more sensitive to violations of quark-hadron duality 
than $x$. Recent claims \cite {LIGETI} that there is a model independant 
estimate yielding $x$, $y$ around 1\% are greatly overstated.

After the early indication from FOCUS that $y$ might be around 1\%, 
other data, also from the $B$ factories have not confirmed it 
\cite{PUROHIT}: $y$ as well as $x$ are consistent with zero on the 
1-2 \% level. Detailed searches for CP asymmetries in $D$ decays 
have been and are being undertaken: data are consistent with no 
asymmetries so far, again on the few percent level \cite{STENSON}.

While it is possible to construct New Physics scenarios producing 
effects as large as 10 \% or so, a more reasonable range is the 
1\% level. Thus one is only now entering territory where there are some 
realistic prospects for New Physics to emerge. 
\end{itemize}

\subsection{The `Novel Challenge'}

The situation is quite different in $B$ transitions since the 
CKM dynamics already generate large CP asymmetries. The one 
significant exception arises in 
$B_s(t) \to \psi + \eta/\phi$ where one can reliably predict a small
asymmetry  not exceeding 2 \% for reasons that are very specific to the 
CKM description \cite{BS} 
\footnote{On the leading level quarks of only the second and third family 
participate.}; 
anything beyond that is a manifestation 
of New Physics.

As presented in the talks by Matteuzzi \cite{MATT} and 
Yamamoto \cite{YAMA}, we can expect that the (Super-)B factories, 
BTeV and LHC-b will allow to measure a host of CP asymmetries 
with experimental uncertainties not exceeding a few percent. 
The question then arises whether we can exploit this 
level of sensitivity {\em theoretically}; i.e., whether one can 
interprete data and make predictions with no more than a 
few percent theoretical uncertainty. I do not consider such a goal for
theoretical  control a luxury. With the exception of 
$B_s(t) \to \psi +\phi/\eta$ noted above CP asymmetries in $B$ decays
often  are large within the SM (or are severely restricted by the 
need for strong phase shifts). Therefore New Physics typically cannot
change 
SM predictions by orders of magnitude. Furthermore I had argued 
above that
the success  of the CKM theory in describing weak observables
characterized by  scales ranging over several orders of magnitude is
highly non-trivial.  Accordingly I do not find it very likely that New
Physics will  affect transition rates for
$B$ hadrons in a {\em massive} way. For to have escaped detection 
before it had to `know' about the flavour structure of the SM. 
While such a feat might be turned by some SUSY implementation 
of New Physics, one cannot count on it.
  
Thus one is faced with a novel challenge: can one be 
confident of having established the presence of New Physics when 
the difference between the expected and the observed signal is much 
less than an order of magnitude? To be more specific: assume 
one predicts an asymmetry of 40 \%, yet observes 60 \% -- can one 
be certain of New Physics? What about if one observes 50\%? 
Interpreting such {\em quantitative} 
discrepancies represents a challenge 
which we have not faced before.

\section{News from the Lepton Sector}
\label{LEP}

\subsection{Charged current interactions}

Important lessons can still be learnt from charged current transitions, 
in particular in $\tau$ decays as discussed by K\"uhn \cite{KUHN}. 
Efforts are under way to extract the strange quark mass $m_s$ and 
$|V(us)|$ from Cabibbo suppressed $\tau$ decays; one gets very 
`reasonable values', however at present the theory uncertainties are 
under poor control. One
can also search for CP asymmetries, which would establish the 
intervention of New Physics, in single $\tau \to K \pi \nu$ decays 
{\em without} $\tau$ polarization.

\subsection{Flavour changing neutral currents}

Lepton flavour changing neutral current transitions represents the 
most intriguing aspects of lepton dynamics. They require  
neutrinos to be mass {\em non}degenerate. Theorists have thought 
hard and long about a reason for neutrinos being massless. With them 
never having succeeded I conclude there is no such reason -- and
therefore  neutrinos are not massless after all. The `see-saw' mechanism 
requiring the existence of right-handed neutrinos and a Majorana 
mass $M_M$ much larger than Dirac fermion masses $m_D$  provides a very 
appealing solution since it implies the existence of very heavy 
mostly right-handed neutrinos and mostly left-handed neutrinos with 
masses $\sim m_D^2/M_M$. With neutrinos not mass-degenerate
lepton-flavour changing transitions  can take place.

The present bound on $\mu \to e \gamma$ reads \cite{KETTLE} 
\beq 
{\rm BR}(\mu \to e \gamma ) \leq 1.2 \cdot 10^{-11} 
\eeq 
The intervention of New Physics is unequivocally needed to 
create a signal that could ever be observed. 
An experiment at PSI aims at a sensitivity level of around 
$10^{-14}$! No miracle is needed for a signal to surface at that 
level.

Searches for $\mu - e$ conversion are complementary. The best available 
bound has been obtained in the quasi-elastic reaction 
\beq 
\frac{\sigma (\mu ^- + Au \to e^- + Au)}
{\sigma (\mu ^- + Au \to \mu ^- + Au)} \leq 6.1 \cdot 10^{-11} \; . 
\eeq  
There is the ambitious goal to go even after the $10^{-17}$ level 
\cite{KETTLE}!

For a long time neutrino oscillations and their consequences have been 
searched for. Finally it seems they have made their presence felt and it
makes even sense to distinguish between different oscillation 
scenarios \cite{FOGLIO}. 
The solar neutrino deficit and the atmospheric muon anomaly 
are widely taken to show clear signals of neutrino oscillations 
\footnote{The LSND signal can be incorporated only at the price of a
sterile neutrino or even CPT violation.}.  
With respect to the solar $\nu$
deficit the evidence is based mainly  and -- after SNO -- robustly on the
total `disappearance' rate with  first indications arising now about more
unique signals like the  energy etc. dependance. As far as the
atmospheric anomaly is  concerned, the deficit in the muon signal is seen
as robust;   more intriguing since more specific is the very robust
azimuthal dependance  of the signal and the evidence for the appropriate
energy  dependance.

A priori there are several contenders for the correct oscillation
scenario. The leading one {\em now} is the L(arge)M(ixing)A(ngle)
case. It used to be quite different, and 
the shift came about due to 
more data, mainly from SNO and K2K. The latter observes 
44 fully contained events when $64 ^{+6.1}_{-6.6}$ are 
expected without oscillation; the {\em no} oscillation 
hypothesis is presently disfavoured at the $\sim 2 \sigma$ 
level \cite{CHAUSSARD}. I just
cannot resist to point out a parallel to the soccer  world championship
going on right now. There was a lot of conviction about who the leading
contenders were; yet this clear picture has been  scrambled significantly
by new data, i.e. the outcome of real  soccer matches! 2002 has the
potential to provide us with some  decisive progress in our knowledge
about the neutrino sector, but --  unlike soccer -- it will not tell us
yet who the true champion is.

\subsection{"If the gods want to harm you ..." -- CP violation 
in the lepton sector}

Above I have emphasized what the CKM theory can do. However it has 
a significant deficit as well: it cannot generate the baryon 
number of the universe. At present it seems likely that baryogenesis 
actually represents a shadow of leptogenesis. To generate a lepton 
number for the universe one needs CP violation in the lepton sector. 
It is clearly desirable to directly study CP violation in leptodynamics. 
This can be done by 
\begin{itemize}
\item 
probing for an electric dipole moment of electrons, atoms 
and molecules, 
\item 
searching for CP asymmetries in $\tau$ decays, 
\item 
analyzing the muon transverse  polarization in $K_{\mu 3}$ decays and
\item 
probing for CP violation in neutrino oscillations. 
\end{itemize} 
As far as
theoretical control is concerned  neutrino physics in general and
oscillations in particular are  optimal systems. However the situation
reminds me of a saying by the  ancient Greek:"If the gods want to really
harm you, then they fulfill your  wishes!" It appears extremely
challenging or iffy if CP violation  in neutrino oscillations could ever
be established because of the neutrino parameters indicated by the
present oscillation phenomenology  and due to matter oscillations
introducing an environmental  bias \cite{CHAUSSARD}. There is one general
lesson to be derived from this: we should be  careful in what we wish and
be thankful for hadronization as explained in the beginning.

\section{Outlook}
\label{OUT}

"The SM is consistent with the data" is a statement most of you 
experience as a worn-out refrain. However in the last two years it has 
acquired new dimensions (pun intended) 
leaving the Higgs sector as the only remaining  
`terra incognita' of the SM. For an essential test of the 
CKM description of CP violation has been performed in $B \to \psi K_S$; 
the first CP asymmetry outside the $K^0 - \bar K^0$ complex has 
been observed, and it is huge -- as expected.  
In my judgement the CKM description of 
CP violation thus has been promoted from an ansatz to a tested theory 
that is going to stay with us. Yet this success of the CKM theory 
does not resolve any of the central mysteries of the SM concerning the 
heavy flavour sector: why is there family replication, why are there 
three families, what generates the very peculiar pattern in the 
quark masses and the CKM parameters? It actually 
deepens those mysteries and -- in my view -- makes a convincing 
case that the SM is incomplete.

This conclusion is further strengthened by three observations: 
\begin{enumerate} 
\item 
One might argue that neutrino oscillations can be 
incorporated into a `trivial' extension of the SM by just 
adding right-handed neutrinos without gauge interactions; one 
can engineer neutrino Yukawa couplings to Higgs doublets 
in such a way as
to  obtain the needed mass matrices. However that would be highly 
contrived; the only known {\em natural} way to understand the 
tiny neutrino masses is, as already stated, 
through the see-saw mechanism,
which  requires Majorana masses. Yet those cannot be obtained from
doublet  Higgs fields! While the see-saw mechanism suggests a highly  
hierarchical structure in the neutrino parameters, this is not
necessarily so as pointed out by Jezabek, since there are 
actually two matrices describing $\nu$ mass-related parameters 
\cite{JEZABEK}.  
\item 
The `strong CP problem' remains unsolved. 
\item 
We know now that CKM dynamics cannot generate 
the baryon number of the Universe.    
\end{enumerate} 
It is obvious then that the dedicated study of heavy flavour dynamics 
can never become marginal, let alone obsolete. 
Much more can be said about this; here I want to comment only on 
directions for CP studies. It hardly needs justification to analyze 
all kinds of CP asymmetries in the decays of beauty hadrons with as 
much precision as possible. Yet this truth should not make us forget 
about other important avenues to pursue. For the 
{\em non}-CKM CP violating dynamics needed to generate the Universe's 
baryon number could well be buried in $B$ decays under the CKM
`background'  of huge CP asymmetries. Telegdi's dictum can be 
applied here in a modified way: 
"yesterday's sensation" -- 
CP violation in $B_d \to \psi K_S$ -- "is today's calibration" 
-- for CP violation in $B_d \to \pi \pi$ -- 
"and tomorrow's background" -- when searching for what is generating 
the baryon number of the Universe. Their
impact on ordinary matter  made up from the light flavours would have to
deal with hardly a  competition from CKM dynamics. At the same time we
would benefit  tremendously from the expertise accumulated and the
opportunities  spotted in different areas: atoms, molecules and nuclei
can  represent promising labs to search for $T$ {\em odd} effects like 
EDM's etc.
 
I have mentioned in the beginning that we have two `Dresden 
codices' to decipher, namely the CKM, 
Eq.(\ref{DRES}, and MNS matrices. Like with 
the original Dresden codex we can succeed only by examining many 
correlations with accuracy and dedication. Gaining much more 
experimental information will be crucial -- yet by itself not 
sufficient. When all the data are `in', we have to do more than 
just connect and interprete them -- we have to understand them. 
That -- in my not quite unbiased view -- means that one will need 
decisive input from theorists!

\section*{Acknowledgments}
Many thanks go to Franco Grancagnolo and his team for organizing an
inspiring meeting in the gorgeous setting of Vietri; re-adjusting 
to normal life without the sweeping views Vietri offered is not easy. 
I am very grateful
to Roberto Perrino for his very generous help with his computer savvy
and intuition. I have learnt a lot from the expertise of the speakers
and  benefitted from further discussions in particular with N. Uraltsev, 
U. Meissner, J. Miranda and M. Neubert.  
This work has been
supported by the NSF under the grant PHY-0087419.


\begin{thebibliography}{99}

\bibitem{BUTLER}
G. Rochester, C. Butler, {\em Nature (London) 160} (1947) 855.

\bibitem{PURCELL} 
E.M. Purcell, N.F. Ramsey, {\em Phys.Rev.} {\bf 78} (1950) 807.

\bibitem{LEO} 
For pictures see: http://www.amnh.org/exhibitions/codex/

\bibitem{DRESDEN} 
For more details see: 
http://www.archaeoastronomie.de/codex/cdstart.htm

\bibitem{PAUL} 
S. Paul, these Proceed.

\bibitem{GOTT}
E. Gottschalk, these Proceed.

\bibitem{MARCH}
J. March-Russell, these Proceed.

\bibitem{MARTINELLI}
G. Martinelli, these Proceed.

\bibitem{DOK} 
I. Bigi {\em et al.}, 
{\em Phys. Lett.} {\bf B 181} (1986) 157.

\bibitem{NASON}
P. Nason, these Proceed.

\bibitem{CDF} 
D. Acosta {\em et al.}, {\em Phys. Rev.} {\bf D65} (2002) 052005.

\bibitem{NASON2} 
M. Cacciari, P. Nason, hep-ph/0204025.

\bibitem{GLAD} 
L. Gladilin, these Proceed.

\bibitem{MIGL}
P. Miggliozzi, these Proceed.

\bibitem{COL} 
G. Colangelo, these Proceed.

\bibitem{ROOS} 
H. Leutwyler, M. Roos, {\em Z. Phys.} {\bf C25} (1984) 91.

\bibitem{CIRI} 
V. Cirigliano {\em et al.}, {\em Eur. Phys. J.} 
{\bf C23} (2002) 121.

\bibitem{DPF02}
A. Sher, talk given at DPF 2002, April 2002.

\bibitem{KLEIN} 
K. Kleinknecht, these Proceed.

\bibitem{ISI} 
G. Isidori, these Proceed.

\bibitem{KETT}
S. Kettell, these Proceed.

\bibitem{WANKE} 
R. Wanke, these Proceed.

\bibitem{DELUCIA} 
E. DeLucia, these Proceed.

\bibitem{BCP} 
I.I. Bigi, in: 
Proc. of `Heavy Quarks at 
Fixed Target 2000', I. Bediaga, J. Miranda, A. Reis (eds.), 
Frascati Physics Series, Vol. XX; 
for a review with references to the
earlier  literature, see: B. Bellini, I. Bigi and 
P. Dornan, {\em Phys. Rep.} {\bf 289} (1997) 1.

\bibitem{GUB}
B. Guberina, H. Stefancic, {\em Phys. Rev.} {\bf D65} (2002) 114004. 
\bibitem{BUCCELLA} 
F. Buccella, these Proceed.

\bibitem{CLEO} 
S. Ahmed {\em et al.}, CLEO Collab., 
{\em Phys. Rev. Lett.} {\bf 87} (2001) 251801.

\bibitem{KOD} 
For an expert discussion, see: A. Khodjamirian, 
{\em AIP Conf.Proc.} {\bf 602} (2001) 194-205, 
hep-ph/0108205.

\bibitem{BIANCO}
FOCUS Collab., J.M. Link {\em et al.}, hep-ex/0206069.

\bibitem{MIRANDA} 
J. Miranda, these Proceed.

\bibitem{MEISS} 
U.-G. Meissner, these Proceed.

\bibitem{MARI} 
N. Marinelli, these proceed.

\bibitem{MARBELLA}
I.I. Bigi, in: Proceed. of the 3rd Workshop on the Tau-Charm 
Factory, Marbella, Spain, 1993.

\bibitem{WANG}
C.-H. Wang, these Proceed.

\bibitem{ROBERTSON}
S. Robertson, these Proceed.

\bibitem{YAMA} 
H. Yamamoto, these Proceed.

\bibitem{URI}
N.G. Uraltsev, these Proceed.; 
preprint hep-ph/0010328, to appear in the B. Joffe Festschrift 
{\em At the Frontier of Particle Physics/Handbook of QCD}, M. Shifman
(ed.), World Scientific, Singapore, 2001.

\bibitem{NEUBERT} 
M. Neubert, these Proceed.


\bibitem{MEL}
K. Melnikov, A. Yelkhovsky, {\em Phys. Rev.} {\bf D 59} 
(1999) 114009.

\bibitem{HOANG} 
A. Hoang, {\em Phys. Rev.} {\bf D 61} (2000) 034005.

\bibitem{SIGNER}
M. Beneke and A. Signer, {\em Phys. Lett.} {\bf B 471} (1999) 233.

\bibitem{ART}
A. Snyder, H. Quinn, {\em Phys. Rev.} {\bf D48} (1993) 2139.

\bibitem{LANDAU}
For a colourful description of Landau's attitude, see 
B.L. Ioffe, preprint hep-ph/0204295.

\bibitem{EPSPRIME} 
S. Giudice, these Proceed.

\bibitem{CS}
A. B. Carter, A.I. Sanda, {\em Phys. Rev.} {\bf D 23} (1981) 1567.

\bibitem{BS}
I.I. Bigi, A.I. Sanda, {\em Nucl. Phys.} {\bf B 193} (1981) 85.

\bibitem{BABAR1}
S. Ricciardi, these Proceed.

\bibitem{BABAR2} 
B. Aubert {\em et al.}, hep-ex/0207042.

\bibitem{YAMA}
M. Yamauchi, Plenary talk at ICHEP'02, Amsterdam, 2002.

\bibitem{EPR} 
A. Einstein, B. Podolsky and N. Rosen, 
{\em Phys. Rev.} {\em 47} (1935) 777.

\bibitem{KRAKOW} 
I. I. Bigi, Invited Talk given at `Meson 2002', Krakow, 
Poland, May 24 - 28, 2002, preprint UND-HEP-02-BIG05.

\bibitem{TRIESTE} 
S. Bertolini, J. Eeg, M. Fabbrichesi, {\em Rev. Mod. 
Phys.} {\bf 72} (2000) 65.

\bibitem{BERT}
S. Bertolini, these Proceed.

\bibitem{BEFORETOP} 
I.I. Bigi, Invited lecture given at the Rencontres de 
Moriond, March 1992, preprint UND-HEP-92-BIG01.

\bibitem{DENT} 
T. Dent, J. Silva-Marcos, hep-ph/0206086.

\bibitem{LACKER} 
F. Parodi, P. Roudeau and A. Stocchi, hep-ph/9903063.

\bibitem{POSP} 
M. Pospelow, these Proceed.

\bibitem{DOSC}
I.I. Bigi, N. Uraltsev, {\em Nucl. Phys.} {\bf B592}(2001) 92;  
for an earlier analysis see: H. Georgi, {\em Phys. Lett.} 
{\bf B297} (1992) 353.

\bibitem{LIGETI}
A.F. Falk, Y. Grossman, Z. Ligeti and A.A. Petrov, 
{\em Phys. Rev.} {\bf D 65} (2002) 054034.

\bibitem{PUROHIT}
M. Purohit, these Proceed.

\bibitem{STENSON}
K. Stenson, these Proceed.


\bibitem{MATT}
C. Matteuzzi, these Proceed.

\bibitem{KUHN}
J. K\"uhn, these Proceed.

\bibitem{KETTLE} 
S. Kettle, these proceed.

\bibitem{FOGLIO}
G. Foglio, these Proceed.

\bibitem{CHAUSSARD}
L. Chaussard, these Proceed.

\bibitem{JEZABEK} 
M. Jezabek, hep-ph/0205234, to appear in Acta 
Physica Polonica B.
























\end{thebibliography}
\end{document}